\def\LCDM{$\Lambda$CDM}
\def\hkpc{$h^{-1}{\ }{\rm kpc}$}
\def\hMpc{$h^{-1}{\ }{\rm Mpc}$}
\def\Msun{${\rm M_{\odot}}$}
\def\hMsun{$h^{-1}{\ }{\rm M_{\odot}}$}

\def\nbody{$N$-body}

\def\Rvir{$R_{\rm vir}$}
\def\Mvir{$M_{\rm vir}$}

\def\zform{$z_{\rm form}$}

\newcommand{\Table}[1]{Table~\ref{#1}}
\newcommand{\Sec}[1]{Section~\ref{#1}}

\newcommand{\Fig}[1]{Figure~\ref{#1}}
\newcommand{\mlapm}{\texttt{MLAPM}}

\def\ea{et~al.~}                            
\def\esyr{events star$^{-1}$ yr$^{-1}$}

\documentclass[useAMS,usenatbib,usegraphicx]{mn2e}
\usepackage[english]{babel}
\title[MACHOs in dark matter haloes]
{MACHOs in dark matter haloes}

\author[Holopainen et al.]
{	Janne Holopainen$^{1}$, Chris Flynn$^{1}$, Alexander Knebe$^{2}$, Stuart P. Gill$^{3}$, Brad K. Gibson$^{4}$ \\
	$^1$Tuorla Observatory, V\"ais\"al\"antie 20, Piikki\"o, FIN-21500, Finland\\
	$^2$Astrophysikalisches Institut Potsdam, An der Sternwarte 16, 14482 Potsdam, Germany \\
	$^3$Columbia University, Department of Astronomy, 550 West 120th Street, New York, NY 10027, USA \\
	$^4$Centre for Astrophysics, University of Central Lancashire, Preston, PR1 2HE, United Kingdom}
	
\begin{document}
\maketitle

\begin{abstract}

Using eight dark matter haloes extracted from fully-self consistent cosmological
$N$-body simulations, we perform microlensing experiments. A hypothetical
observer is placed at a distance of 8.5 kpc from the centre of the halo measuring optical depths, event
durations and event rates towards the direction of the Large Magellanic
Cloud. We simulate 1600 microlensing experiments for each halo.  Assuming that
the whole halo consists of MACHOs, $f = 1.0$, and a single MACHO mass is
$m_{\rm M} = 1.0$ \Msun, the simulations yield mean values of $\tau = 4.7^{+5.0}_{-2.2} \times 10^{-7}$
and $\Gamma = 1.6^{+1.3}_{-0.6} \times 10^{-6}$ \esyr.  We find
that triaxiality and substructure can have major effects on the measured values
so that $\tau$ and $\Gamma$ values of up to three times the mean can be found.
If we fit our values of $\tau$ and $\Gamma$ to the MACHO
collaboration observations \citep{macho}, we find $f =
0.23^{+0.15}_{-0.13}$ and $m_{\rm M} = 0.44^{+0.24}_{-0.16}$. Five out of the
eight haloes under investigation produce $f$ and $m_{\rm M}$ values mainly
concentrated within these bounds.

\end{abstract}

\section{Introduction}

Experiments such as MACHO \citep{macho} and POINT-AGAPE \citep{point-agape} have detected microlensing 
events towards the Large Magellanic Cloud (LMC) and the M31,
supporting the view that some fraction of the Milky Way dark halo may be comprised
of massive astronomical compact halo objects (MACHOs).
The fraction of the dark halo mass in MACHOs, as well as the nature of
these lensing objects and their individual masses can be constrained to some
extent by combining observations of microlensing events with analytical models
of the dark halo (e.g. \citealt{kerins}; \citealt{macho}; \citealt{cardone}).
However, the real dark halo might not be as simple as in analytical models,
which typically have spherical or azimuthal symmetry and, in particular,
smoothly distributed matter. Cosmological simulations indicate that dark matter
haloes are {\it neither} isotropic nor homogeneous but are rather triaxial
(e.g. \citealt{warren}) and contain a notable amount of substructure
(e.g. \citealt{klypin99}). Even for the Milky Way it is still unclear if the
enveloping dark matter halo has spherical or triaxial morphology: a detailed
investigation of the tidal tail of the Sgr dwarf galaxy supports the notion of a
nearly spherical Galactic potential (\citealt{ibata}; \citealt{majewski}) whereas that the data
are also claimed to be consistent with a prolate or oblate halo \citep{helmi04}.
The shape of the halo has an effect on the predicted number of microlenses
along different lines-of-sight, for comparison with the results of microlensing
experiments.

In this paper, we examine the effects of dark halo morphology and dark halo
clumpiness on microlensing surveys conducted by hypothetical observers located
in eight $N$-body dark matter haloes formed fully self-consistently in
cosmological simulations of the concordance \LCDM\ model. The eight haloes are
scaled to approximately match the Milky Way's dark halo in mass and rotation
curve properties, and hypothetical observers are placed on the surface of a ``Solar sphere''
(i.e. 8.5 kpc from the dark halo's centre) from where they conduct microlensing
experiments along lines-of-sight simulating the Sun's line-of-sight to the LMC.
The effects on microlensing of triaxiality and clumping in the haloes are examined.

The paper expands upon work in a study by \cite{wd}
although we take a slightly different point of view. Instead of studying the errors caused by
different analytical models fitted to an $N$-body halo, we analyse the
microlensing survey properties of the simulated haloes directly. That is, we do not
try to gauge the credibility of analytical microlensing descriptions but rather
use our self-consistent haloes for the \textit{inverse problem in microlensing}
\citep{cardone}. In other words, the free parameters (the MACHO mass and
the fraction of matter in MACHOs) are determined from the observations via the
optical depth, event duration and event rate predictions of the models.  This
not only allows us to make predictions for these parameters based upon fully
self-consistent halo models but also to investigate the importance of shape and
substructure content of the haloes.

The paper is structured as follows. We describe the $N$-body haloes and their
preparation in Section 2, the simulation details are covered in Section 3,
equations for the observables are derived in Section 4, results are given in
Section 5, some discussion is presented in Section 6, final conclusions are in Section 7.

\section{The Haloes}

\subsection{Cosmological simulation details}

Our analysis is based on a suite of eight high-resolution \nbody\ simulations
\citep{gillA} carried out using the publicly available adaptive
mesh refinement code \mlapm\ \citep{knebe} in a standard \LCDM\
cosmology ($\Omega_0 = 0.3,\Omega_\lambda = 0.7, \Omega_b h^2 = 0.04, h = 0.7,
\sigma_8 = 0.9$). Each run focuses on the formation and evolution of 
galaxy cluster sized object containing of order one million collisionless, dark
matter particles, with mass resolution $1.6 \times 10^8$ \hMsun\ and force
resolution $\sim$2\hkpc\ (of order 0.05\% of the host's virial radius). The
simulations have sufficient resolution to follow the orbits of satellites
within the very central regions of the host potential ($\geq$ 5--10 \% of the
virial radius) and the time resolution to resolve the satellite orbits with
good accuracy (snapshots are stored with a temporal spacing of $\Delta t
\approx$ 170 Myr).  Such temporal resolution provides of order 10-20
timesteps per orbit per satellite galaxy, thus allowing these simulations to be
used in a previous paper to accurately measure the orbital parameters of each
individual satellite galaxy \citep{gillB}.

The clusters were chosen to sample a variety of environments. We define the
virial radius $R_{\rm vir}$ as the point where the density of the host
(measured in terms of the cosmological background density $\rho_b$) drops below
the virial overdensity $\Delta_{\rm vir}=340$. This choice for $\Delta_{\rm
vir}$ is based upon the dissipationless spherical top-hat collapse model and is
a function of both cosmological model and time.  We further applied a lower
mass cut for all the satellite galaxies of $1.6 \times 10^{10}$ \hMsun\ (100
particles).  Further specific details of the host haloes, such as masses,
density profiles, triaxialities, environment and merger histories, can be found
in \cite{gillA} and \cite{gillB}. \Table{HaloDetails}
gives a summary of a number of relevant global properties of our halo
sample. There is a prominent spread not only in the number of satellites with
mass $M_{\rm sat} > 1.6 \times 10^{10}$\hMsun\ but also in age, reflecting the
different dynamical states of the systems under consideration.

\begin{table}
\caption{
	Summary of the eight host dark matter haloes. The superscript $cl$
	indicates that the values are for the unscaled clusters.  Column 2
	shows the virial radius, $R_{\rm vir}$; Column 3 the virial mass,
	$M_{\rm vir}$; Column 4 the redshift of formation, \zform; Column 5 the
	age in Gyr; and the final column an estimate of the number of
	satellites (subhaloes) in each halo, $N_{\rm sat}$.}
\begin{tabular}{cccccc}\hline
  Halo 
& \Rvir$^{cl}$ 
& \Mvir$^{cl}$
& \zform 
& age 
& $N_{\rm sat}$ \\
& \hMpc
& $10^{14}$ \hMsun
& 
& Gyr
& \\
 
\hline \hline
 \# 1 &  1.34 & 2.87 & 1.16 & 8.30 & 158 \\
 \# 2 &  1.06 & 1.42 & 0.96 & 7.55 &  63 \\
 \# 3 &  1.08 & 1.48 & 0.87 & 7.16 &  87 \\
 \# 4 &  0.98 & 1.10 & 0.85 & 7.07 &  57 \\
 \# 5 &  1.35 & 2.91 & 0.65 & 6.01 & 175 \\
 \# 6 &  1.05 & 1.37 & 0.65 & 6.01 &  85 \\
 \# 7 &  1.01 & 1.21 & 0.43 & 4.52 &  59 \\
 \# 8 &  1.38 & 3.08 & 0.30 & 3.42 & 251 \\
\hline
\label{HaloDetails}
\end{tabular}
\end{table}

\subsection{Downscaling}
As we intend to perform microlensing experiments directly comparable to the
results of the MACHO collaboration we need to scale our haloes to the size of
the Milky Way. For this purpose we follow \cite{helmi03} and
apply an adjustment to the length scale, requiring that densities remain
unchanged. Hence the scaling relations are:

\begin{eqnarray}
r & = & \gamma r^{cl} \\
v & = & \gamma v^{cl} \\
m & = & \gamma^3 m^{cl} \\
t & = & t^{cl}
\end{eqnarray}

\noindent
where $r$ is any distance, $v$ is velocity, $m$ is mass, $t$ is time and the
superscript $cl$ refers to the unscaled value. Because the haloes are
downscaled, $\gamma$ will always be in the range \hbox{$\gamma \in [0,1]$}.

\begin{figure}
\includegraphics[height=84mm, angle=-90]{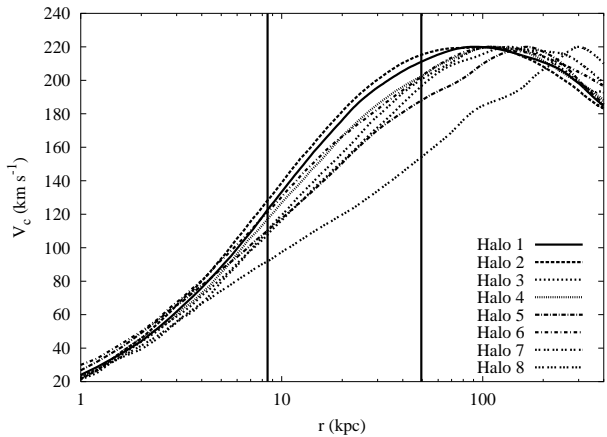}
\caption{
	Circular velocity curves of the downscaled haloes ($V_c =
	\sqrt{GM(r)/r}$).  Note that the maximum of each curve is at \hbox{220
	km\,s$^{-1}$.} The scaling factors $\gamma$ are shown in
	\Table{ScaledHaloes}.  The two vertical lines at 8.5 kpc and 50.1 kpc
	mark the distances of the Sun and the LMC from the centre of the Milky
	Way.}
\label{vcprofiles}
\end{figure}

To find a suitable $\gamma$, the maximum circular velocity of each halo is
required to be 220 km\,s$^{-1}$. The resulting (scaled) rotation curves for all
eight haloes are shown in \Fig{vcprofiles}. This figure highlights that Halo
\#8 is a special system --- its evolution is dominated by the interaction of three
merging haloes \citep{gillB}. We do not consider Halo \#8 to be an acceptable
model of the Milky Way, but we include it into the analysis as an extreme case.

One might argue that our (initially) cluster sized objects should not be used
as models for the Milky Way for they formed in a different kind of environment
with less time to settle to (dynamical) equilibrium (i.e. the oldest of our
systems is 8.3 Gyr vs. $\sim$ 12 Gyr for the Milky Way). However, \cite{helmi03}
showed that more than 90\% of the total mass in the central region of a
realistic Milky Way model was in place about 1.5 Gyr after the formation of the
object. Our study focuses exclusively on this central region (i.e. the inner
15\% in radius) boosting our confidence that our haloes do serve as credible
models of the Milky Way for our purposes.

In \Table{ScaledHaloes} we list the scaled radii and masses, as well as the
scaling factor $\gamma$, for each halo. Note that we have adopted a Hubble
constant of $h = 0.7$, which applies throughout the paper. The circular
velocity (rotation) curves for the scaled haloes are shown in \Fig{vcprofiles}
and demonstrate how the scaled mass $M(<r)$ is accumulated out to 400 kpc.  Two
vertical lines mark the position of observers at the Solar circle and the
distance of the LMC from the halo centre; within this region, we note that the
mass profiles are similar in all haloes, however, there are differences,
especially for halo~\#8 (the dynamically merging system of haloes).

\begin{table}
\centering
\caption{
	Properties of the haloes after downscaling to Milky Way size.  Column 2
	shows the scale factor for each halo, $\gamma$; Column 3 shows the
	scaled virial radius, $R_{\rm vir}$; Column 4 the scaled virial mass,
	$M_{\rm vir}$; and the final column the scaled particle mass, $m_p$.}
\begin{tabular}{lcccc}
\hline
 Halo & $\gamma$ & $R_{\rm vir}$ & $M_{\rm vir}$       & $m_p$ \\ 
      &          & kpc           & $10^{12}$ M$_\odot$ & $10^{6}$ M$_\odot$ \\
\hline
\hline
\#1 & 0.197 & 380 & 3.08 & 1.75 \\
\#2 & 0.248 & 378 & 3.00 & 3.47 \\
\#3 & 0.253 & 391 & 3.39 & 3.71 \\
\#4 & 0.277 & 388 & 3.23 & 4.86 \\
\#5 & 0.197 & 383 & 3.14 & 1.76 \\
\#6 & 0.267 & 402 & 3.61 & 4.33 \\
\#7 & 0.278 & 403 & 3.67 & 4.92 \\
\#8 & 0.213 & 421 & 4.19 & 2.21 \\
\hline
\label{ScaledHaloes}
\end{tabular}
\end{table}

\subsection{Characteristics}

Dark matter haloes within cosmological simulations are very different from the idealised analytical
dark haloes that are described by only a few parameters. Two significant characteristics of the dark
haloes that are not captured by the analytical models are triaxialty and substructure.
For example, within the simulations on both cluster and galactic scales, the dark matter
haloes are triaxial rather than spherical (\citealt{warren}, \citealt{jing}).
However, when gas dynamics is included within the simulation, these clusters/galaxies are
more spherical (\citealt{katz}, \citealt{evrard}, \citealt{dubinski}), and when cooling
is included within the simulation, the haloes are even more spherical \citep{kaza}.

Simulations show that CDM haloes contain a large number of subhaloes at both galactic
and cluster scales (i.e. satellites) (\citealt{klypin99}, \citealt{moore}),
further \cite{moore} assert that galaxy haloes appear as scaled versions of galaxy
clusters with respects to the subhalo populations.

However, within the inner regions of the host haloes (on both galactic and cluster scales)
the number density of subhaloes decreases considerably (\citealt{ghigna}, \citealt{gillB}, \citealt{diemand}).
Even though these results seem to hold under numerical convergence tests \citep{diemand},
the dynamical properties of the dark matter subhaloes differ from the observed galaxies within clusters.

With the use of semi-analytical techniques these differences have been
accounted for (\citealt{gao}; \citealt{taylor}). \cite{gao} conclude that the
tidal stripping of the dark haloes was very efficient in reducing the dark
matter mass of the subhaloes while having less of an effect on the tightly bound
baryonic galaxy at its core. Thus, while the dark matter subhaloes in the inner
regions were being disrupted, baryonic cores survived.

For our pure dark matter simulations \citep{gillB} we also see this lack of
dark matter substructure in the inner regions. This is shown in \Table{Nsat},
where we list the subhaloes (along with some of their (downscaled) integral
properties) found in the inner 70 kpc of each halo. Note that masses of the
subhaloes are limited to a relevant range, $5.0 \times 10^{8}$ \Msun $< M_{\rm
sat} < 1.0 \times 10^{10}$ \Msun.

For the Milky Way, we know that at least one large satellite exists within the
inner 70 kpc: the LMC. Thus, whatever the paucity of satellites in inner
regions of dark haloes in the simulations, here at least is one real-world
example. The mass of the LMC is estimated to be $M_{\rm LMC}(8.9 {\rm~kpc}) =
8.7 \pm 4.3 \times 10^{9}$ \Msun\ from kinematical data \citep{marel}.
Reassuringly enough, we do find a few satellites with similar masses
(cf. \Table{Nsat}) in the simulated haloes. Whether there are more (dark)
satellites in the inner regions of the Milky Way than in our models is an open
question.

\begin{table}
\centering
\caption{
	Substructure within the inner 70 kpc of the (downscaled) haloes.  Column
	1 shows the name of the halo; Column 2 the distance from the centre of
	the subhalo to the centre of the host, $l_{\rm sat}$; Column 3 the
	truncation radius of the subhalo (within the truncation radius, the
	density of the subhalo is larger than the background density of the
	host halo) , $r_{\rm sat}$; Column 4 the number of bound particles in
	the subhalo, $N_{\rm sat}$; and the final column the mass of the
	subhalo, $M_{\rm sat}$.  }
\label{subhaloes}
\begin{tabular}{ccccc}
\hline
Halo & $l_{\rm sat}$ & $r_{\rm sat}$ & $N_{\rm sat}$ & $M_{\rm sat}$ \\
& kpc & kpc & particles & $10^{9}$\Msun \\
\hline
\hline
\#1 & 67.0 & 34.6 & 1232 & 2.20 \\
\#1 & 20.0 & 45.4 & 2957 & 5.27 \\
\#1 & 23.5 & 27.9 & 632 & 1.13 \\
\\
\#2 & 53.5 & 24.1 & 207 & 0.73 \\
\#2 & 44.8 & 23.3 & 186 & 0.66 \\
\\
\#3 & 60.7 & 38.1 & 793 & 2.99 \\
\#3 & 63.0 & 32.7 & 466 & 1.76 \\
\\
\#4 & 67.8 & 30.8 & 317 & 1.57 \\
\\
\#7 & 38.5 & 48.7 & 1217 & 6.09 \\
\#7 & 62.4 & 22.6 & 124 & 0.62 \\
\\
\#8 & 46.1 & 32.9 & 859 & 1.93 \\
\hline
\label{Nsat}
\end{tabular}
\end{table}

\section{Setting the Stage}

\subsection{The microlensing cone}

A Galactocentric coordinate system is given by the three eigenvectors of the
inertia tensor of the respective halo. We choose the $z$-axis to coincide with
the major axis. To conduct our microlensing experiments, we then place an
observer at the Solar distance from the centre of a halo.  There are various
options for possible orientations with respect to the coordinate system; we
basically restrict ourselves to the scenario where the observer is placed
somewhere on a sphere of radius 8.5 kpc.

\begin{figure}
\includegraphics[width=84mm, angle=0]{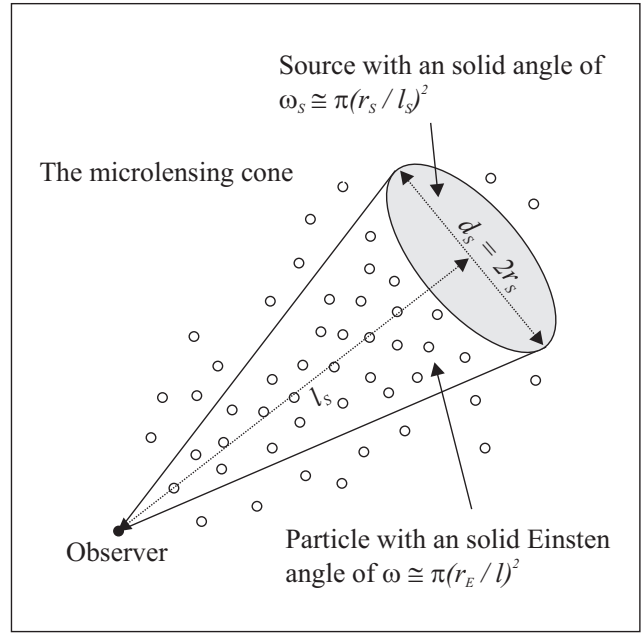}
\caption{
	The microlensing sightline is defined by the locations of the observer
	and the source and the cone by the sightline and the source diameter.
	We use $l_S = 50.1$ kpc and $d_S = 4.52$ kpc. The particles inside the
	cone are treated as gravitational lenses. Typically \hbox{$\sim$ 100}
	particles from the cosmological simulation are found inside a
	cone. This number is increased to $\sim$ 10000 by breaking them up into
	subparticles according to the recipe outlined in Section
	\ref{subparticles}.  }
\label{microlensingcone}
\end{figure}

We sample the halo with a number of sightlines and combine the individual
results obtained from these independent measurements to obtain values of
interest.  Once we put down an observer, we can construct an observational cone
towards a region simulating observations of an LMC sized patch of sky, and
compute the microlensing signal generated by particles in the cone out to a
distance of 50.1 kpc (see Figure \ref{microlensingcone}). The hypothetical LMC
has also to be located in a realistic direction in respect to the observer.

\begin{figure}
\includegraphics[width=84mm, angle=0]{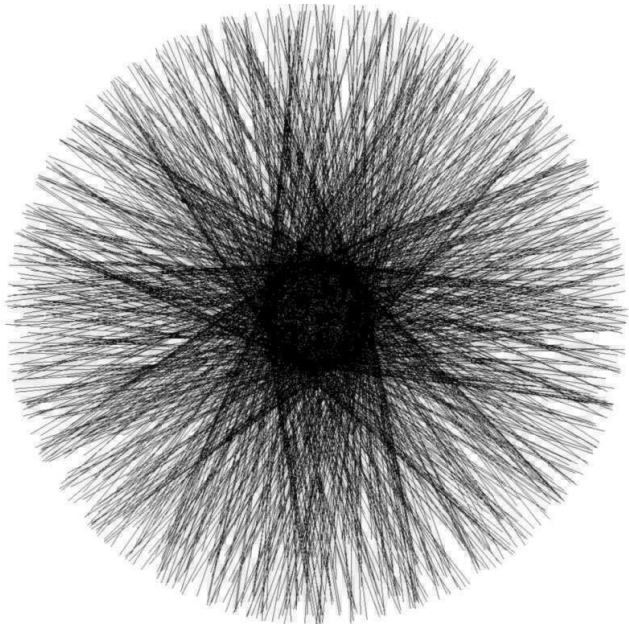}
\caption{
	Sightlines used in the microlensing simulation. The image shows all
	1600 sightlines of the 40 observers.  The observers are located on a
	sphere with a radius of $r_\odot = 8.5$ kpc. The sphere can be seen as
	the ``dark ball'' in the centre of the sightlines. The radius of the
	sphere which holds the sources is 49.5 kpc.  Every observer has 40
	sources (LMCs) at a distance of 50.1 kpc.  The sightlines are not
	uniformly distributed because there is a limited number of observers
	and the angle $(\vec{r}_{\rm obs},~\vec{r}_{LMC} - \vec{r}_{\rm obs})$
	is kept fixed at 89.1$^\circ$.  This is the original angle between
	$(\vec{r}_\odot,~\vec{r}_{LMC} - \vec{r}_\odot)$ in Galactocentric
	coordinates.  }
\label{cones}
\end{figure}

We use 40 observers uniformly distributed on a sphere of radius $8.5$ kpc.
Each observer observes 40 uniformly separated LMCs, for which the angle
$(\vec{r}_{\rm obs},~\vec{r}_{LMC} - \vec{r}_{\rm obs})$ is kept fixed at
89.1$^\circ$ (the original angle between $(\vec{r}_\odot,~\vec{r}_{LMC} -
\vec{r}_\odot)$ in Galactocentric coordinates).  This configuration
(illustrated in Figure \ref{cones}) gives 1600 individual sightlines, and
microlensing cones, which sample different sets of particles within a halo. The
choice for the number of sightlines is a compromise between computation time
and sampling density. As seen in Figure \ref{cones}, 1600 sightlines sample the
volume quite sufficiently, even when the sightlines are drawn as lines instead
of actual cones.

The microlensing cone itself is defined by the observer and the
source. \Fig{microlensingcone} illustrates that we treat the source (i.e. LMC)
as a circular disk with diameter $d_S$. The particles inside the cone are
considered to be gravitational lenses. However, this kind of setup can lead to
a handful of particles (those closest to the observer) contributing {\it most
to the optical depth}. This leads to high sampling noise, and is essentially
due to the limited resolution of the cosmological simulation (see Figure
\ref{ODComparison}). It follows that the particle distribution has to be
smoothed in some manner to suppress this noise; we describe this and the
related issue of the MACHO masses in the next section.

\subsection{Mass resolution vs. MACHO mass} \label{subparticles}

In order to compute microlensing optical depths and other properties from the
particles in the simulations, we need to get from a mass scale of order $10^6$
\Msun, i.e. that of the particles in the simulation, down to the mass scale of
the MACHOS, of order 1 \Msun; i.e. we account for the fact that a typical
particle in the simulations represents of order $10^6$ MACHOs.

To overcome the limitation imposed by the mass resolution of the simulation we
recall that individual particles in the simulation are not treated as
$\delta$-functions but have a {\it finite size}. The extent of a particle
(i.e. its size) is determined --- in our case --- by the spacing of the grid,
as we are using an adaptive mesh refinement code (i.e. \mlapm). In \mlapm\ the
mass of each particle is assigned to the grid via the so-called
triangular-shaped cloud (TSC) mass-assignment scheme \citep{hockney}
which spreads every individual particle mass over the host and surrounding $3^3
- 1$ cells. The corresponding particle shape (in 1D) reads as follows:

\begin{equation} \label{TSCloud}
 S(x) = \left\{ \begin{array}{ll}
                  \displaystyle 1-\frac{|x|}{L} & \mbox{for $|x| < L$}\\
                  0               & \mbox{otherwise}
                \end{array}
        \right.
\end{equation}

\noindent

where $L$ is the spacing of the grid and $x$ measures the distance to the
centre of the cell.

For each particle present in and around a particular cone along an observer's
line-of-sight, we determine the size of the finest grid surrounding it, which
in turn determines the physical extent of the particle.  The particle is then
re-sampled with 100 ``subparticles'' whose positions are randomly chosen under
the density distribution given by Equation \ref{TSCloud}.  Figure
\ref{boxes} illustrates these ``subparticle clouds'', sampled according to this 
TSC mass-assignment.

\begin{figure}
\includegraphics[height=84mm, angle=-90]{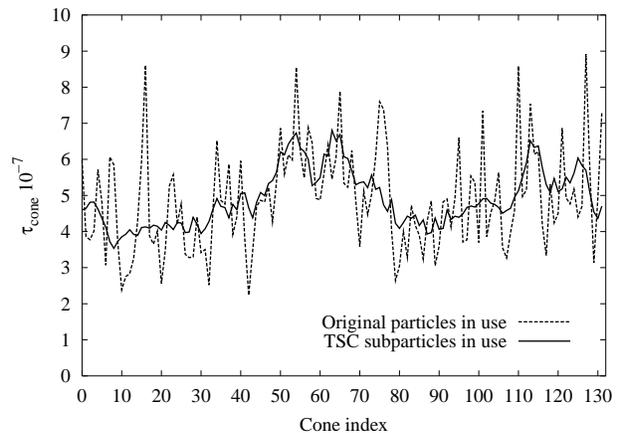}
\caption{
	Comparison between the use of cosmological simulation particles and
	triangular-shaped clouds containing 100 subparticles. For the
	illustration, we distributed 132 observers on a perimeter of a disk
	with a radius of 8.5 kpc, located on the $xy$-plane. The volumes of consecutive cones
	overlap by half.  Without any resolution enhancements the variation in
	optical depth from cone to cone can be as large as a factor of three!
	This is due to the insufficient mass resolution of the cosmological
	simulation. We reached the Poissonian noise level, in which cone-to-cone
	variations are $< 20$ \%, with 100 subparticles, by experimenting. The
	large scale trend is due to triaxiality of the halo.}
\label{ODComparison}
\end{figure}

\begin{figure}

\includegraphics[width=84mm, angle=0]{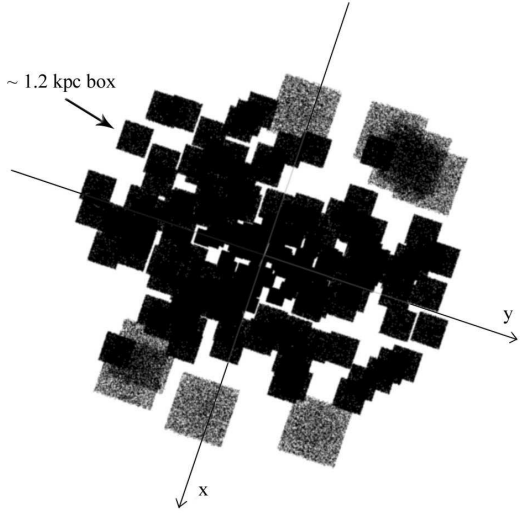}
\caption{
	For this illustrative example of the subparticle clouds, we placed 222
	cosmological simulation particles onto the $xy$-plane and broke each
	particle down to 10,000 subparticles.  To get the subparticle positions,
	we used the triangular-shaped cloud density profile identically to the
	actual cosmological simulation. The subparticles form a cube (of a size
	of the corresponding \mlapm\ grid cell) around the position of the
	cosmological simulation particle.  In the real microlensing simulation,
	the whole volume under investigation is filled with intersecting cubes
	and the void areas between the cubes seen here are not present.  }
\label{boxes}

\end{figure}

The improvement gained by using this method can be viewed in \Fig{ODComparison}
where we show measured optical depths (to be defined in \Sec{OpticalDepths}) in
a series of microlensing cones towards the LMC for observers continuously
rotated on the edge of a disc perpendicular to the major axis of the dark
halo. What is shown here is the noisy ``original'' estimates of optical depth
for many lines-of-sight, compared to the optical depth measurements obtained
when the particles within the cones have been resampled to ``subparticles''.
In the un-resampled case, the handful of particles which happen to reside close
to the observer are found to dominate the calculation of the optical depth and
other quantities of interest such as event duration and event rate. Variations
in the optical depth from sightline to sightline can vary by up to a factor
of three in the un-resampled sample cones, simply because a handful of
particles dominate the microlensing. Convergence experiments have shown that
breaking the original simulation particles down to 100 subparticles is
sufficient to reduce the noise from this source to be negligible; if we were to
use more subparticles, there is no further gain in accuracy.

\begin{figure}

\includegraphics[width=84mm, angle=0]{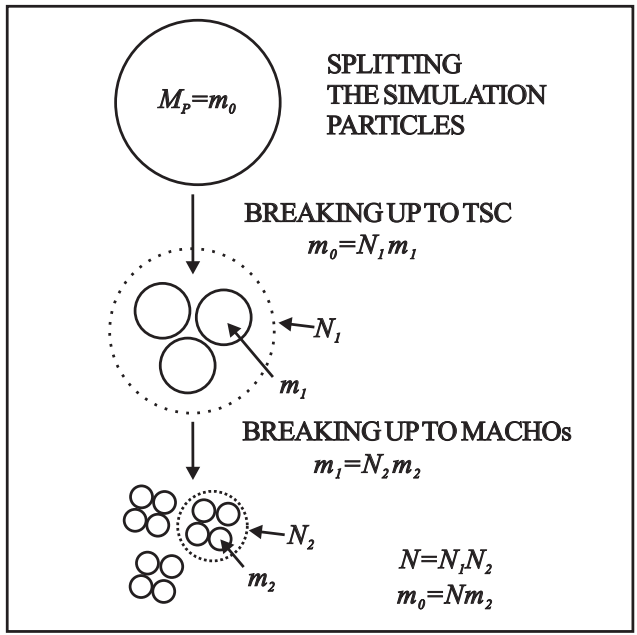}
\caption{
	Splitting the cosmological simulation particle to 100 subparticles and
	a subparticle to $N_{\rm M}$ MACHOs.  In the first phase, we break the
	cosmological simulation particle to 100 subparticles.  The resulting
	subparticle represents an order of $N_{\rm M} \sim 10^4$ MACHOs with a
	mass of 1 M$_\odot$.  The subparticle is treated as a MACHO
	concentration where the represented MACHOs all have a single location
	($l_{\rm M} \equiv l_{\rm sub}$), mass ($m_{\rm M} \equiv m_{\rm sub} /
	N_{\rm M}$) and velocity ($v_{\rm M} \equiv v_{\rm sub}$).  }
\label{particlesplitting}

\end{figure}

The subparticles introduced have masses of order 10$^{4}$ M$_\odot$, so we are
unfortunately still orders of magnitudes away from actual MACHO masses.  In the
formulae introduced later (\Sec{Experiments}), each subparticle is expressed as
a singular concentration of MACHOs. That is, a subparticle is treated as a set
of MACHOs which have the same position and velocity as the subparticle itself,
and for convenience in the simulations the MACHO mass is chosen to be $m_{\rm M} =
1.0$ \Msun (although this is later relaxed by scaling). \Fig{particlesplitting}
demonstrates the hierarchy of particles starting from the cosmological
simulation particle down to the individual MACHOs.

\section{The Microlensing Experiments} \label{Experiments}

\subsection{Definitions}
Before going to the microlensing equations, we define what we mean by certain
terms:

{\bf A microlensing source} is a circular region, oriented perpendicularly to the
line-of-sight of the observer, in which {\it uniformly distributed} background
source stars are located.  Because of the statistical nature of the source
model, the number of background stars affects only the microlensing event rate.
The region with a diameter $d_S$ and a distance $l_S$, is shown in
\Fig{microlensingcone}.

{\bf A source star} is a star located somewhere in the disk of the microlensing
source.

{\bf A microlens} is also a circular region, oriented perpendicularly to the
line-of-sight of the observer and centered on a dark particle. The region has a
radius $r_{\rm E}$, which depends on the location and mass of the
particle. Lenses are always inside a microlensing cone between the source and
the observer.

{\bf A microlensing event} is a detectable amplification of a source star
caused by a microlens. Detectable means that the light of a source star is
amplified by a factor larger than 1.34. This occurs when the sightline to a
source star passes a particle within the lens radius, also known as the
Einstein radius.

\subsection{Optical depth} \label{OpticalDepths}

The Einstein radius $r_{\rm E}$ of a gravitational lens is defined as

\begin{equation} \label{EinsteinRadius}
r_{\rm E}^2 = \frac{4Gm}{c^2} l^2 \biggr( \frac{1}{l} - \frac{1}{l_S}\biggr) =
A m l^2 \biggr( \frac{1}{l} - \frac{1}{l_S} \biggr),
\end{equation}

where $m$ is the mass of the lens, $l$ is the distance to the lens, $l_S$ is
the distance to the source and $A = \frac{4G}{c^2}$.

For a single subparticle, we define the lens radius as

\begin{equation} \label{r_sub}
r_{\rm sub}^2 = A m_{\rm sub} l_{\rm sub}^2 \biggr( \frac{1}{l_{\rm sub}} -
\frac{1}{l_S} \biggr),
\end{equation}

where $m_{\rm sub}$ is the mass of the subparticle and $l_{\rm sub}$ the
distance between the subpartice and the observer.  Note that all subparticles
within a halo have the same mass.

The optical depth of a single lens describes the probability that any given
source star is amplified by the lens at any given time. Thus, the general
equation for the optical depth of a lens can be written as the ratio of the
solid angles of the lens and the source

\begin{equation}
\tau = \frac{\omega}{\omega_S},
\end{equation}

where $\omega$ is the solid Einstein angle of the lens and $\omega_S$ is the
solid angle of the source. When the solid angles are small, this can be
approximated to

\begin{equation}
\tau \simeq \biggr( \frac{r_{\rm E}}{l} \frac{l_S}{r_S} \biggr)^2 
= A m \Theta^{-2} \biggr( \frac{1}{l} - \frac{1}{l_S} \biggr),
\end{equation}

where $r_S$ is the radius of the source and $\Theta^{-2} = (l_S / r_S)^2$. Note
that $\Theta$ is approximately the half opening angle of the cone.

The optical depth of a subparticle with a mass $m_{\rm sub}$ is defined as

\begin{equation} \label{tau_sub}
\tau_{\rm sub} = A \Theta^{-2} m_{\rm sub} \biggr( \frac{1}{l_{\rm sub}} 
- \frac{1}{l_S} \biggr).
\end{equation} 

We take this value to be the total optical depth from all the $\sim 10^4$
MACHOs the subparticle represents.

Optical depth is additive, as long as the lenses do not overlap or cover the
whole source, and so the total optical depth in a microlensing cone which
contains $N$ subparticles is

\begin{equation}
\label{tau_cone}
\tau_{\rm cone} 
	= \sum_{i}^{N} \tau_{\rm sub}^{(i)} = A \Theta^{-2} m_{\rm sub}
	\sum_{i}^{N} \biggr( \frac{1}{l_{\rm sub}^{(i)}} - \frac{1}{l_S}
	\biggr)
\end{equation}

From Equation \ref{tau_cone} it follows that the optical depth in a cone
depends mainly on the distances of the subparticles related to the
observer. The closer the particles are the larger is the optical depth. Section
\ref{subparticles} covered the details of particle breaking, which also has a
large effect on $\tau_{\rm cone}$. Note that $\tau_{\rm cone}$ does not depend
on the MACHO mass $m_{\rm M}$.

\subsection{Event duration}

The event duration of a lens describes the typical duration of the amplifying
event the lens would produce. The detected event durations in the MACHO
experiment are the order of 100 days. Event duration depends on the tangential
velocity with which a lens would seem to pass a source star and on the Einstein
radius of the lens. The equation for an individual subparticle is

\begin{equation} \label{t_sub}
t_{\rm sub} = \frac{\pi}{2} \frac{r_{\rm sub}}{v_{\rm sub}},
\end{equation}

where $r_{\rm sub}$ is the lens radius of the subparticle and $v_{\rm sub}$ the
apparent tangential velocity difference between the subparticle and the source,
respective to the observer.

The term $\frac{\pi}{2}$ is the average crossing length of a circle with an
unit radius.  By adding this term, we correct the event duration from the
maximum value ($2r_{\rm E}/v$) to the statistically expected value.  The
correction can also be seen in the observational optical depth, $\tau_{meas}$,
as a $\pi/4$ term in Eq. 1 of Alcock~\ea (2000).

Using Equation \ref{r_sub}, Equation \ref{t_sub} becomes

\begin{equation}
\label{t_sub2}
  t_{\rm sub} = \frac{\pi}{2} m_{\rm sub}^{1/2} \frac{l_{\rm sub}}{v_{\rm sub}}
  \biggr[ A \biggr(\frac{1}{l_{\rm sub}} - \frac{1}{l_S} \biggr) \biggr]^{1/2}.
\end{equation}

However, this is not yet the value we are after. We need to solve $\hat{t}_{\rm
sub}$, which is the average event duration caused by the subparticle's
MACHOs. To get $\hat{t}_{\rm sub}$, we simply replace $m_{\rm sub}$ with
$m_{\rm M}$ and assign $l_{\rm M} \equiv l_{\rm sub}$ and $v_{\rm M} \equiv
v_{\rm sub}$. We get

\begin{equation}
\label{t_sub3}
 \hat{t}_{\rm sub} \equiv t_{\rm M} \equiv \frac{\pi}{2} m_{\rm M}^{1/2}
 \frac{l_{\rm sub}}{v_{\rm sub}} \biggr[ A \biggr(\frac{1}{l_{\rm sub}} -
 \frac{1}{l_S} \biggr) \biggr]^{1/2},
\end{equation}

where $t_{\rm M}$ is the event duration of a single MACHO when the MACHO
inherits the location and the velocity of the subparticle. What is essentially
stated in Equation \ref{t_sub3} is that we use the event duration of a single
MACHO as the average event duration for the whole subparticle.

As is well known, Equation \ref{t_sub2} shows that, unlike the optical depth,
event duration is a function of the MACHO mass. This fact can be used to find a
preferred MACHO mass for a given event duration.

\subsection{Event rate}
Event rate $\Gamma$ for a given experiment is simply the number of expected
events per an observing period for a given source.  For a single lens it can be
given as $\Gamma = \tau / t$, i.e., as the ratio of the optical depth and event
duration.

The MACHO collaboration observed $\sim$ 13 - 17 events in 5.7 years. Usually,
$\Gamma$ is given in events star$^{-1}$ yr$^{-1}$. The MACHO collaboration had
$10.7 \times 10^{6}$ source stars which gives (a maximum) of $\Gamma = 17 / 5.7
/ (10.7 \times 10^{6})$ events star$^{-1}$ yr$^{-1} = 2.79
\times 10^{-7}$ events star$^{-1}$ yr$^{-1}$.

We calculate the event rate for a subparticle consisting of $N_{\rm M}$ MACHOs as
\begin{equation} \label{gamma}
\Gamma_{\rm sub}
	= N_{\rm M} \Gamma_{\rm M} 
	= N_{\rm M} \frac{\tau_{\rm M}}{t_{\rm M}} 
	= N_{\rm M} \frac{\tau_{\rm sub} / N_{\rm M}}{\hat{t}_{\rm sub}} 
	= \frac{\tau_{\rm sub}}{\hat{t}_{\rm sub}},
\end{equation}

where $\Gamma_{\rm M} = \frac{\tau_{\rm M}}{t_{\rm M}}$ is the event rate for a
single MACHO. By using Equations \ref{tau_sub} and \ref{t_sub2}, the last term
of Equation \ref{gamma} expands to

\begin{equation} \label{gamma2}
 \Gamma_{\rm sub} = \frac{2}{\pi} A^{1/2} \Theta^{-2} m_{\rm M}^{-1/2} m_{\rm
 sub} \frac{v_{\rm sub}}{l_{\rm sub}} \biggr(\frac{1}{l_{\rm sub}} -
 \frac{1}{l_S}\biggr)^{1/2}.
\end{equation}

For the cone, we simply sum the event rates of the subparticles together, as
was done to the optical depths in Equation \ref {tau_cone}

\begin{equation}
\label{gamma_cone}
\Gamma_{\rm cone} 
	= \frac{2}{\pi} A^{1/2} \Theta^{-2} m_{\rm M}^{-1/2} m_{\rm sub}
	\sum_{i}^{N} \frac{v_{\rm sub}^{(i)}}{l_{\rm sub}^{(i)}}
	\biggr(\frac{1}{l_{\rm sub}^{(i)}} - \frac{1}{l_S}\biggr)^{1/2}.
\end{equation}

Note that $\Gamma_{\rm sub} \propto m_{\rm M}^{-1/2}$ whereas $\hat{t}_{\rm
sub} \propto m_{\rm M}^{1/2}$.
 
\section{Results}
\subsection{Differential optical depths and event rates}

The results of our simulations are compared to the observations via the event
duration because this is the only {\it directly} observed quantity. Our aim is
to determine how optical depth $\tau$ and event rate $\Gamma$ depend on event
duration.

In Figure \ref{sphere_cones}, we show $d\tau/d\hat{t}$ and $d\Gamma/d\hat{t}$,
averaged over all the 1600 cones for each halo, i.e we show for each simulated
halo the global behaviour of differential optical depths and event rates as a
function of event duration. The normalised standard deviations in these
quantities amongst sightlines are shown (in separate panels, for
clarity). Both curves for $\tau$ and $\Gamma$ peak in the range 50 and 100
days. This was expected as it is the result obtained from analytical models
(e.g. \citealt{kerins}). Our simulations and the analytical results are in this
sense adequate fits to the actual observations, in which the (few) events
cluster around 100 days duration. 

In Table \ref{ResTab}, we list the modal values
for the distributions, $\hat{t}_{\rm exp}$, and the averaged cone values,
$\langle\tau_{\rm cone}\rangle$ and $\langle\Gamma_{\rm cone}\rangle$, which are numerically integrated values of the differential functions in Figure \ref{sphere_cones}.
Note that the error limits on the last row (labelled ``mean'') are simple mean values of the error limits from the corresponding column because they represent the mean variations between cones in a halo, not variations between haloes.      

It can be seen that the integrated values differ between haloes. One might argue that there is a trend
towards lower values from Halo \#1 to Halo \#8, and this is perhaps no surprise
because the haloes were ordered by their dynamical age. Similar grouping of
curves can be seen in the circular velocity profiles in Figure
\ref{vcprofiles}. Because circular velocity profiles reflect density profiles,
it is the differences in matter densities between haloes that can be seen to
cause differences in the integrated values of $d\tau/d\hat{t}$ and
$d\Gamma/d\hat{t}$.

We have applied the MACHO collaboration efficiency functions to our simulations
in order to get an idea of how their efficiency affects the overall
observables. The results can be seen in Tables \ref{ResTabA} and
\ref{ResTabB}. The efficiency functions reduce $\langle\tau_{\rm cone}\rangle$
and $\langle\Gamma_{\rm cone}\rangle$ by more than a factor of two and shift
the expected event durations to be longer.

Note that all the following results are given in terms MACHO's efficiency
function ``A'' --- this is the more conservative option of two efficiency
functions given by the MACHO collaboration; tests show that both give very
similar results.

\begin{table}
\centering
\caption{
	The mean optical depths and event rates, with their standard
	deviations, over the 1600 cones for each halo.  See Figure
	\ref{sphere_cones} for the corresponding (differential) distribution
	functions --- the numbers represented here are the integrated values of
	those functions. We also list the modes of the distributions:
	$\hat{t}_{\rm exp}$ columns give the most probable event duration for
	the respective distribution. Error limits show the variations between cones 
	(i.e. microlensing experiments) in a halo.}
\begin{tabular}{lcccc}
\hline
 Halo & $\langle\tau_{\rm cone}\rangle$    & $\hat{t}_{\rm exp}$ & $\langle\Gamma_{\rm cone}\rangle$  & $\hat{t}_{\rm exp}$ \\ 
      & $10^{-7}$		 & days       & $10^{-6}$        & days     \\
      & & & events & \\
      & & & star$^{-1}$ yr$^{-1}$ & \\
\hline
\hline
\#1 & $5.6^{+6.0}_{-2.7}$ & 84 & $2.0^{+1.7}_{-0.8}$ & 66 \\
\#2 & $5.9^{+7.5}_{-2.7}$ & 74 & $2.1^{+2.1}_{-0.8}$ & 64 \\
\#3 & $4.3^{+3.8}_{-1.8}$ & 84 & $1.4^{+1.1}_{-0.5}$ & 64 \\
\#4 & $5.1^{+5.6}_{-2.4}$ & 86 & $1.6^{+1.3}_{-0.7}$ & 64 \\
\#5 & $4.1^{+3.4}_{-2.2}$ & 91 & $1.3^{+0.9}_{-0.6}$ & 69 \\
\#6 & $5.0^{+4.9}_{-1.9}$ & 81 & $1.7^{+1.3}_{-0.6}$ & 71 \\
\#7 & $4.7^{+6.6}_{-2.5}$ & 84 & $1.6^{+1.6}_{-0.7}$ & 61 \\
\#8 & $2.6^{+2.1}_{-1.0}$ & 104 & $0.8^{+0.6}_{-0.3}$ & 81 \\
\hline
mean & $4.7^{+5.0}_{-2.1}$ & 86 & $1.6^{+1.3}_{-0.6}$ & 68 \\
\hline
\label{ResTab}
\end{tabular}
\end{table}
\subsection{Substructure} \label{res_sub}

An interesting feature can be seen in the relative standard
deviation (i.e. normalised to the global mean) in Figure \ref{sphere_cones} of both $d\tau/d\hat{t}$
and $d\Gamma/d\hat{t}$ for Halo \#1. The relative standard deviation of both
functions has a peak around $\hat{t} = 50$ days of greater than $1.0$ compared to typical
values of $\sim$ 0.7. This peak is due to a subhalo with a mass of $5.27
\times 10^9$ \Msun, i.e. the second subhalo of Halo~\#1 listed in Table 
\ref{subhaloes}. It is the only satellite with a substantial mass, concentration 
and location to cause any features to the $\tau$ and $\Gamma$
distributions. For example, Halo \#7 contains a subhalo with a comparable mass
($6.09 \times 10^9$ \Msun), but because it is located further away from the
centre of the halo (and from the observer), it fails to affect the overall
$\tau$ and $\Gamma$ distributions. The subhalo in Halo \#1 is therefore the
only substructure which could be directly associated with microlensing signal not associated
with the global properties of the haloes.

\begin{figure}

\includegraphics[width=84mm,angle=0]{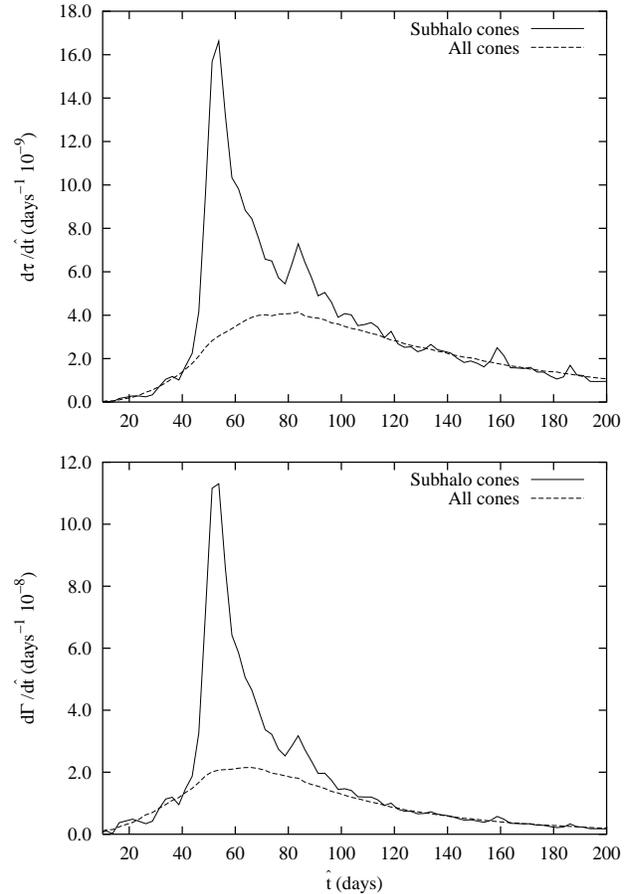}
\caption{
	``Subhalo cones'' shows the average differential functions from 24
	cones which penetrate the second subhalo in Halo \#1 (see Table
	\ref{Nsat}).  ``All cones'' is shown for comparison --- it is the the
	same function as in Figure \ref{sphere_cones} for Halo \#1. The large
	deviation from the average shape is seen in the standard deviation
	panels of Figure \ref{sphere_cones} and here as a sharp rise of both
	$d\tau/d\hat{t}$ and $d\Gamma/d\hat{t}$ functions starting from
	$\hat{t} \sim 40$ days. Individual ``subhalo cones'' produce even more
	distorted distributions and we had to average over the 24 cones to get
	the peak to stand out clearly.  }
\label{clump_diff}

\end{figure}

It should be emphasized that the distributions shown in Figure
\ref{sphere_cones} are {\it averaged} over all cones. The subhaloes, if present
in a halo, contribute only to a small number of the cones and thus, possibly
peculiar distributions are lost by the averaging process. In another words, the
distribution functions of a single cone can be quite different from the
averaged one. A particularly severe case of this can be seen in Figure
\ref{clump_diff} where we show the average distribution functions of 24 cones
which penetrate the second subhalo in Halo \#1 compared to the overall
distributions. The cones are chosen so that their lines-of-sight are closer
than 5 kpc to the centre of the subhalo and therefore the cones contain
particles from the subhalo's core. The cones yield $\tau$ and $\Gamma$ values
which are approximately twice as large as the average values in Table
\ref{ResTab}. Moreover, the event duration peaks ($\hat{t}_{\rm exp}$) are
different from the Table \ref{ResTab} values because the subhalo's particles
have different velocities than the particles of the host halo.

We also probed the cones which penetrate the first subhalo in Halo \#7 for
signatures in the differential functions.  As mentioned earlier, this subhalo
has a comparable mass to the major subhalo of Halo \#1. We found that an excess
signal is also present in the ``subhalo cones'' of Halo \#7. However, this
subhalo does not produce such a peaked functions as seen in Figure
\ref{clump_diff} but instead, $d\tau/d\hat{t}$ and $d\Gamma/d\hat{t}$ stay
steadily above the average shape, at two times the average value,
between $40$ days $<\hat{t} < 200$ days. This roughly doubles the values
of $\tau$ and $\Gamma$ for cones penetrating the subhalo, compared to the average
cone values.

\subsection{Triaxiality}

In addition to substructure, triaxiality was the other halo component which we
wanted to study from the microlensing point of view. Triaxiality effects are
not seen in the $d\tau/d\hat{t}$ and $d\Gamma/d\hat{t}$ functions because they
are averaged over all the cones, and the cones are distributed spherically
inside a halo. However, when the cones are not distributed spherically, one can
immediately see effects due to triaxial shape. Figure \ref{ODComparison} shows
how an observer on a ``Solar orbit'' observes a roughly sinusoidal variation for
the optical depth as function of position on the perimeter. We find the
amplitude to range from $\tau_{\rm min} \sim 4 \times 10^{-7}$ to $\tau_{\rm
max} \sim 6 \times 10^{-7}$ and a peak-to-peak wavelength of $\sim 60$ cones
(which corresponds to 180 degrees in observer positions). This shape is solely
due to triaxiality and is seen in all the haloes.

To measure triaxiality related variations more quantitatively we constructed
Figures \ref{triax_tau} and \ref{triax_gamma}. In both these figures, we have
measured the angle between a positive coordinate axis and the source, keeping in mind that 
the coordinate axes are aligned with the triaxial axes.
After this is done for all cones,
we bin the observables and calculate the upper and lower fractiles so that they
hold 65 \% of the values. Especially the optical depth fractiles reveal clear
triaxiality signal.

In the first panel in Figure \ref{triax_tau}, the mean optical depth is largest
when the sources are close to the $z$-axis (i.e. the major triaxial axis), as
one would expect. For haloes \#1, \#2 and \#7, the optical depth on opposite
sides of the $xy$-plane differs somewhat, demonstrating that matter is not
necessarily distributed with azimuthal symmetry in the simulated (or real) haloes.

There are high mean optical depth values near the positive $y$-axis for Halo
\#1 in the second panel. This anomaly is due to the subhalo discussed in previous Section
\ref{res_sub}. Now we can see that the subhalo is located close to the positive
$y$-axis. This is confirmed by the true location of the subhalo, which is found
to be $\vec{l}_{\rm sat} = (-3.37, 19.8, 1.36)$ kpc.

The lowest panel is different from the two previous ones because we are
measuring {\it tri}axiality. The optical depth values from sources near the
$x$-axis seem to be the lowest amongst all panels. These low values ($\tau_{\rm
cone} \sim 4 \times 10^{-7}$) can also be seen in the second panel as the lower
fractiles at $\alpha \sim 90^\circ$. Thus, by these ``observations'' we can verify that
the $x$-axis is the minor axis. In the lowest panel, the fractiles bend upwards at $\alpha \sim 90^\circ$
because the sources near both the $z$- and $y$-axis produce larger values than the sources near the $x$-axis.

\begin{figure}
\includegraphics[height=84mm,angle=-90]{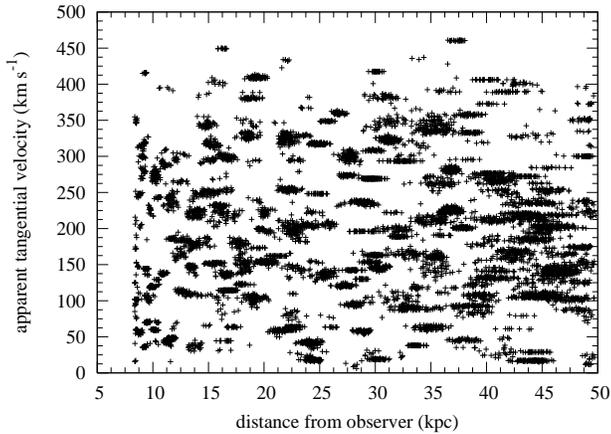}
\caption{
	An example which shows the anticorrelation between apparent tangential velocities and
	locations of subparticles inside a microlensing cone. This anticorrelation explains why
	triaxiality does not show up in $\Gamma$ as clearly as in $\tau$. 
	The size range of the triangular-shaped clouds also shows up clearly in the plot.
}
\label{vl_space}
\end{figure}

\begin{figure}
\begin{center}
\includegraphics[width=150mm]{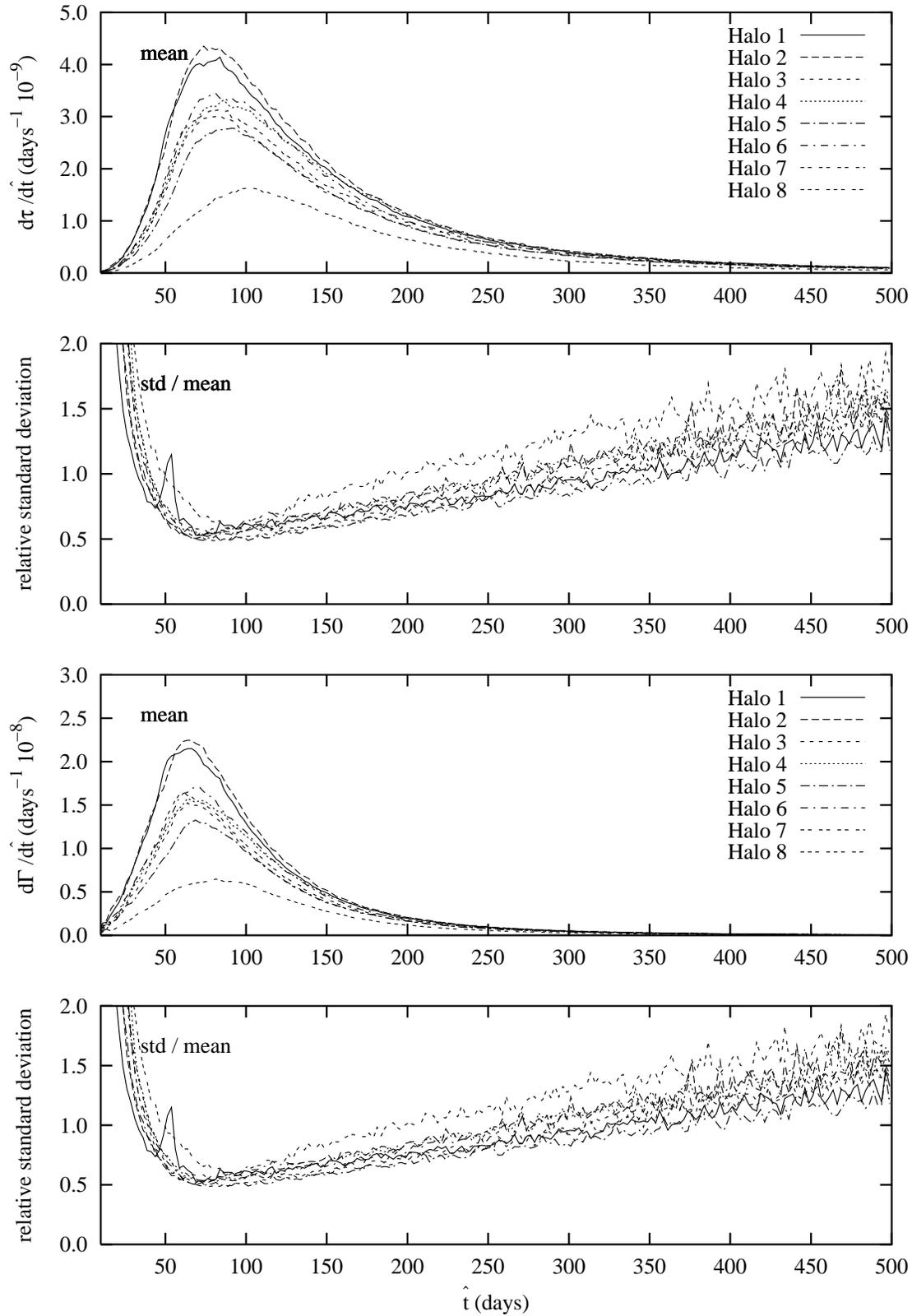}
\caption{
	Differential $\tau$ and $\Gamma$ values as a function of event duration. 
	The largest Eigen-value axis of the halo is along the z-axis,
	and the observers (40) are distributed uniformly onto the ``solar sphere''.
	Because each observer has 40 sources, the mean and the standard deviation are calculated
	from a total of $40 \times 40 = 1600$ cones. This is done by binning the subparticle $\tau$ 
	and $\Gamma$ values for each cone into 200 event duration bins and then calculating 
	the mean and the standard deviation for each bin. Standard deviations are large due to triaxiality which
	causes the main differences in the mass and mass and velocity distributions between sightlines (cones). 
	However, notice the Halo \#1 standard deviation peak. It is due to a subhalo with a mass of $5.27 \times 10^9$ 	\Msun\ (see text and Table \ref{subhaloes}).}
\label{sphere_cones}
\end{center}
\end{figure}

\begin{figure}
\begin{center}
\includegraphics[width=150mm]{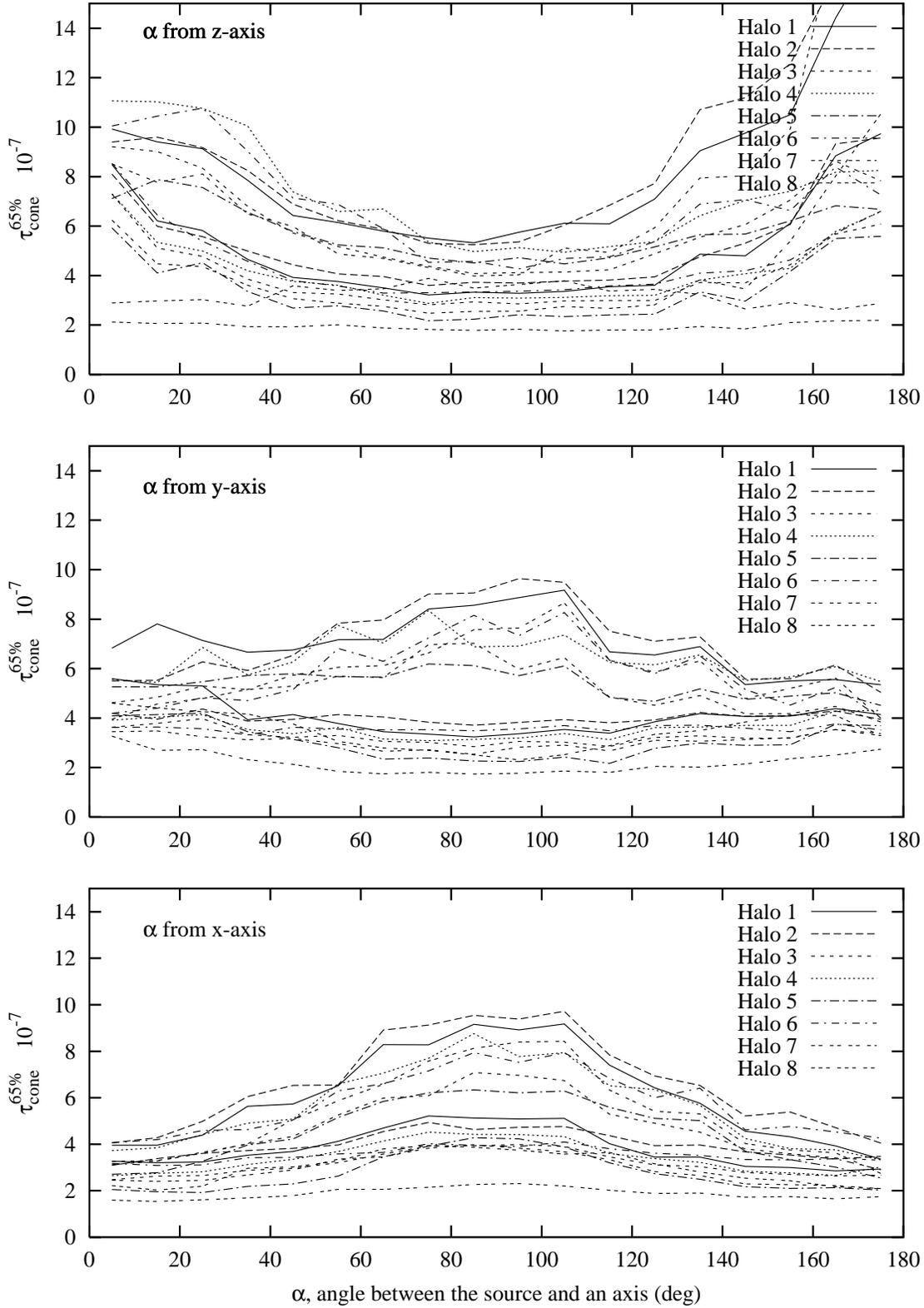}
\caption{
	The clear effect of triaxiality shown by $\tau_{\rm cone}$ fractile boundaries which include 65 \% of the values. The 	$\tau_{\rm cone}$ values are binned in 25 bins by $\alpha$, which is the angle between the source and one of the coordinate 	axes. Fractiles are calculated from $\tau_{\rm cone}$ values within each bin. The major Eigen-value axis of the halo is 	aligned with the z-axis. Notice how $\tau_{\rm cone}$ values grow when the source gets closer to the z-axis.}
\label{triax_tau}
\end{center}
\end{figure}

\begin{figure}
\begin{center}
\includegraphics[width=150mm]{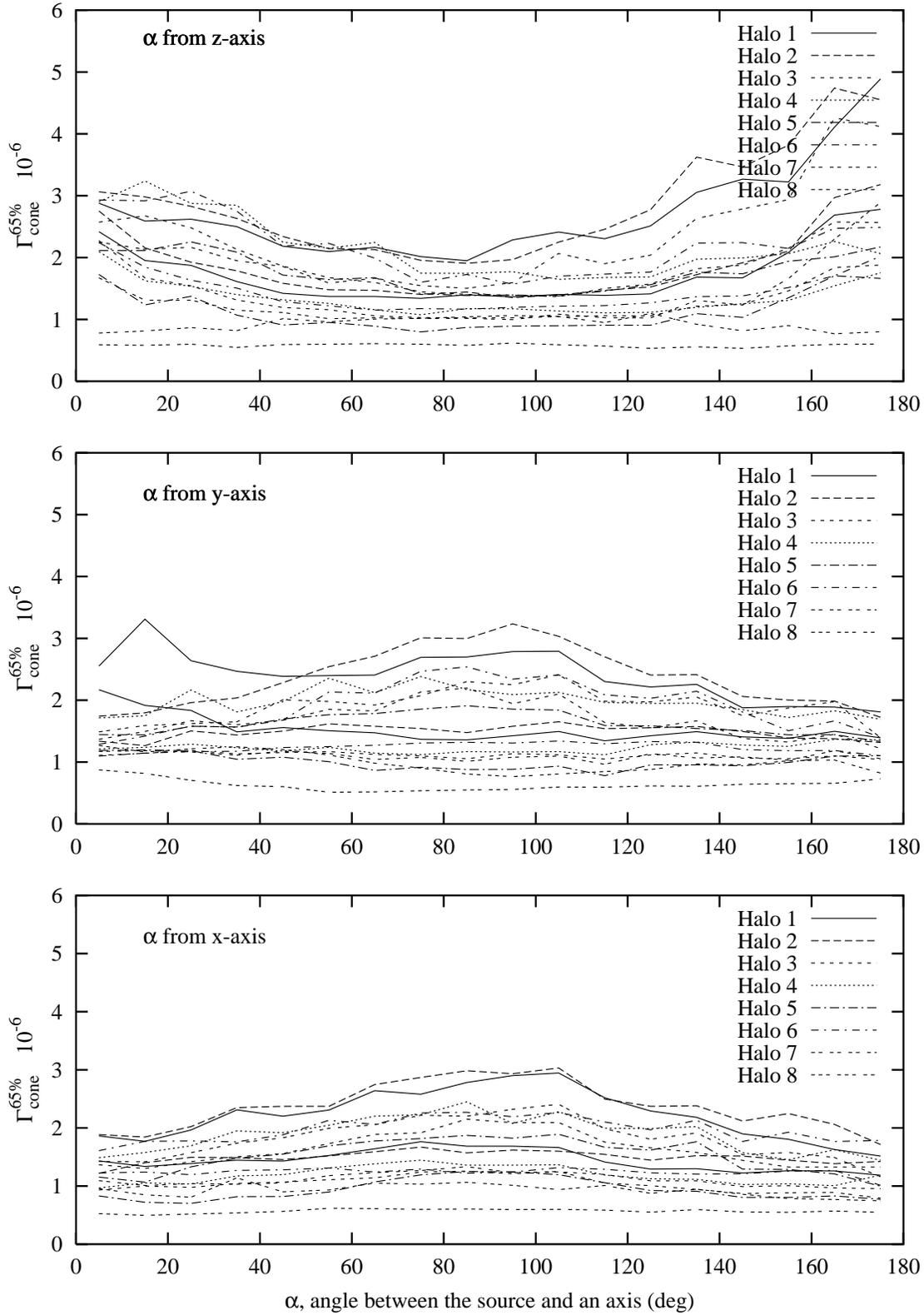}
\caption{
	The same as Figure \ref{triax_tau} but with $\Gamma$ instead of $\tau$.
}
\label{triax_gamma}
\end{center}
\end{figure}

Triaxiality can not be seen as clearly in Figure \ref{triax_gamma} as in Figure
\ref{triax_tau} because $\Gamma$ is affected by the lens velocities whereas
$\tau$ is not. Apparently, the velocities do not carry enough information about
the triaxiality and the correlation between the source location and the
observables are somewhat ``washed out''. This is confirmed
in Figure \ref{vl_space} where we show that there is virtually no correlation
between the apparent tangential velocity and the location of the subparticles
inside a cone. Nevertheless, the most striking
features, i.e. the $z$-axis majority and the subhalo in Halo
\#1, are still clearly visible in $\Gamma$. For example, a suitable subhalo can produce
roughly twice as many events as a similar area without one, based on the second
panel in Figure \ref{triax_gamma}.

\subsection{Estimating MACHO mass and halo mass fraction in MACHOs}

The MACHO collaboration used several analytical halo models to estimate the
typical, individual MACHO mass, $m_{\rm M}$, and the mass fraction of MACHOs in
the halo, $f$, responsible for the lensing events observed. For example, one of
their models fits $m_{\rm M} = 0.6^{+0.28}_{-0.20}$ \Msun\ and $f =
0.21^{+0.10}_{-0.07}$. We calculate our own estimates for $m_{\rm M}$ and $f$
to see how our $N$-body halo models compare to the analytical ones in making
these predictions.

As the observational constraint, we use the blending corrected event durations chosen
by the MACHO collaboration's criterion ``A'', $\hat{t}_{\rm st}(A)$, from Table 8
in Alcock et al. (2000). These values are not corrected for the observational
efficiency function ${\mathcal{E}}(\hat{t})$ nor have they been reduced for the
events caused by the known stellar foreground populations, given in Table 12 in Alcock et
al. (2000). We adopt $N_{\rm exp}(A) = 2.67$ for known population events and
arrive at the observational values of $\tau_{\rm obs} = 3.38 \times 10^{-8}$
and $\Gamma_{\rm obs} = 1.69 \times 10^{-7}$ events star$^{-1}$ yr$^{-1}$.
We then compare these values with our simulation produced values, $\tau_{\rm
cone}$ and $\Gamma_{\rm cone}$, to get $m_{\rm M}$ and $f$. The simulation
values are corrected with the efficiency function ${\mathcal{E}}(\hat{t})$
prior to calculating the predicted values. The calculations are based on the
following equations.

First, we assume that

\begin{equation} \label{mfeq_f}
f \propto \tau,
\end{equation}

which simply describes that that the probability of observing an event is
proportional to the number of lenses in the whole halo. From Equation \ref{gamma_cone},
and from the fact that the number of observed events also follows the total
number of lenses, we get\footnote{Note that in Equation \ref{gamma_cone} we
assume that the whole halo consists of MACHOs, i.e., $f = 1.0$.}

\begin{equation} \label{mfeq_gamma}
\Gamma \propto f m_{\rm M}^{-1/2}.
\end{equation}

These two equations permit us to estimate $m_{\rm M}$ and $f$.  From Equation
\ref{mfeq_f}, we get

\begin{equation}
\frac{f_{\rm obs}}{f_{\rm cone}} = \frac{\tau_{\rm obs}}{\tau_{\rm cone}} \ .
\end{equation}

For the microlensing simulations we assumed $f_{\rm cone} = 1.0$ and thus,

\begin{equation}
f = f_{\rm obs} = \frac{\tau_{\rm obs}}{\tau_{\rm cone}}. 
\end{equation}

This estimate of $f$ is calculated for each cone.

For $m_{\rm M}$, we need to use the event rate in addition to the optical
depth. From Equations \ref{mfeq_f} and \ref{mfeq_gamma}, we get

\begin{equation}
\frac{m_{\rm M}^{obs}}{m_{\rm M}^{cone}} = 
 \biggr(\frac{f_{\rm obs}}{f_{\rm cone}} \frac{\Gamma_{\rm cone}} {\Gamma_{\rm
 obs}}\biggr)^2,
\end{equation}

where $f_{\rm cone} = 1.0$ and $m_{\rm M}^{cone} = 1.0$ \Msun. Thus, we get the
equation for the MACHO mass,

\begin{equation}
 m_{\rm M} = m_{\rm M}^{obs} = \biggr(f_{\rm obs}\frac{\Gamma_{\rm cone}}
 {\Gamma_{\rm obs}}\biggr)^2 {\rm M}_\odot.
\end{equation}

\begin{figure}
\begin{center}
\includegraphics[height=200mm]{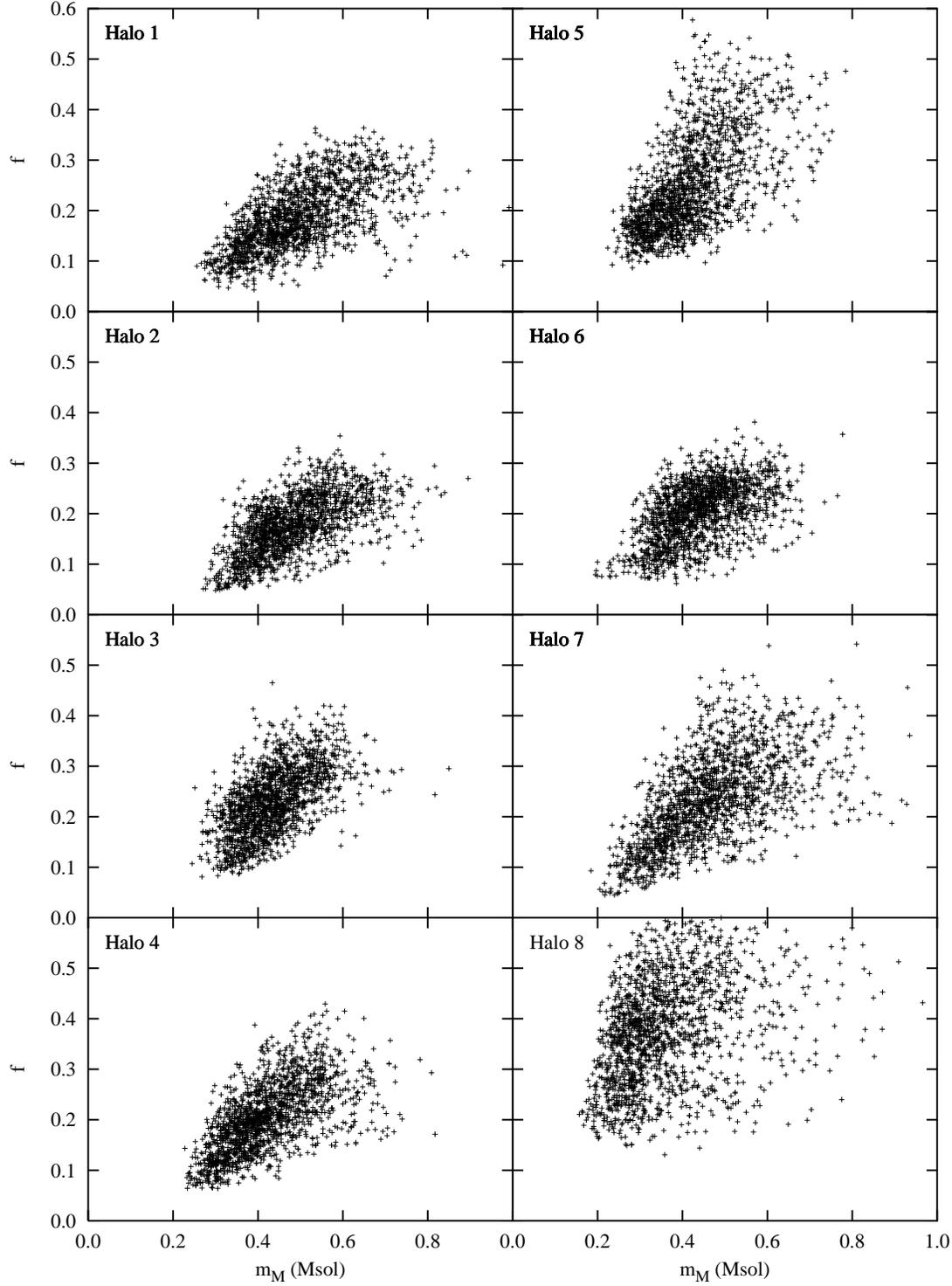}
\caption{
	The predicted MACHO mass ($m_{\rm M}$) and MACHO halo fraction (f) values calculated for each cone.
	The predicted values are calculated as follows:
	$f = \tau_{\rm obs}/\tau_{\rm cone}$ and $m_{\rm M} = (f\Gamma_{\rm cone}/\Gamma_{\rm obs})^2$.
	We derive the observational values from the observed event durations (criteria A, corrected only for blending,
	known populations $N_{\rm exp}(A) = 2.67$ subtracted).
	This leads to $\tau_{\rm obs} = 3.38 \times 10^{-8}$ and $\Gamma_{\rm obs} = 1.69 \times 10^{-7}$ events star$^{-1}$ yr$^{-1}$.
	The efficiency function is applied to the simulation values instead of the observed event durations.
	Our predicted values of $f$ and $m_{\rm M}$ are similar to the MACHO collaboration predictions. However, note the
	differences in the scatter area shapes between haloes. These differences are due to the fact that the haloes have different 	triaxial shapes. Moreover, substructure can cause a small number of individual data points to cluster together.
}
\label{mf_scatter}
\end{center}
\end{figure}

Figure \ref{mf_scatter} shows our computation for $f$ and $m_{\rm M}$ for every
cone in all eight host haloes. The mean values are
$f = 0.23^{+0.15}_{-0.13}$ and $m_{\rm M} = 0.44^{+0.24}_{-0.16}$, where the error limits
show the mean scatter within a halo and contain 95 \% of the values.

Our values populate approximately the same areas
as the contours in Figure 12 of \cite{macho} with the exception that
our mass predictions do not extend to $m_{\rm M} = 1.0$ \Msun\ but stay mainly below
$m_{\rm M} = 0.6$ \Msun. Figure \ref{mf_scatter} therefore can be interpreted
as (another) confirmation that downscaled cluster sized CDM haloes acquired from
cosmological $N$-body simulations can be used in interpreting microlensing
observations as well as analytical models --- at least if the analysis is limited
to the innermost regions.

Note that the scatter in $f$ is solely due to the variations in optical depth in
different experiments, and thus, Figure \ref{mf_scatter} shows directly how optical depth
varies between cones and even haloes. The scatter in $m_{\rm M}$ contains additionally
the variations in event rates between experiments. 

In the calculations above, we have assumed that $\tau_{\rm obs} = 3.38 \times 10^{-8}$ is solely due to MACHOs in the dark halo of the Milky Way. However, if some of the observed events originate from elsewhere (e.g. the observations contain LMC self-lensing), we have overestimated $\tau_{\rm obs}$. This leads to an overestimation of both $f$ and $m_{\rm M}$, and in this sense, the values in Figure \ref{mf_scatter} are upper limits.

\begin{table}
\centering
\caption{
	Same as Table \ref{ResTab} but here we used the MACHO collaboration's
	efficiency function A.  These numbers should correspond to the
	${\mathcal{E}}(\hat{t})$--{\it uncorrected} observations if the MACHO
	fraction would satisfy $f = 1.0$ and the MACHO mass $m_{\rm M} = 1.0$
	\Msun. The distributions are not shown because they are essentially
	similar to Figure \ref{sphere_cones} except for the smaller integrated
	values (given in Tables \ref{ResTabA} and \ref{ResTabB}).}
\begin{tabular}{lcccc}
\hline
 Halo & $\langle\tau_{\rm cone}\rangle$   & $\hat{t}_{\rm exp}$ & $\langle\Gamma_{\rm cone}\rangle$  & $\hat{t}_{\rm exp}$ \\ 
      & $10^{-7}$		 & days       & $10^{-6}$        & days     \\
      & & & events & \\
      & & & star$^{-1}$ yr$^{-1}$ & \\
\hline
\hline
\#1 & $2.1^{+2.2}_{-1.0}$ & 84 & $0.7^{+0.6}_{-0.3}$ & 66 \\
\#2 & $2.2^{+2.8}_{-1.0}$ & 84 & $0.7^{+0.8}_{-0.3}$ & 71 \\
\#3 & $1.6^{+1.4}_{-0.7}$ & 84 & $0.5^{+0.4}_{-0.2}$ & 74 \\
\#4 & $1.9^{+2.1}_{-0.9}$ & 86 & $0.6^{+0.5}_{-0.3}$ & 74 \\
\#5 & $1.5^{+1.3}_{-0.8}$ & 94 & $0.5^{+0.3}_{-0.2}$ & 76 \\
\#6 & $1.8^{+1.9}_{-0.7}$ & 99 & $0.6^{+0.5}_{-0.2}$ & 71 \\
\#7 & $1.7^{+2.5}_{-0.9}$ & 89 & $0.6^{+0.6}_{-0.3}$ & 61 \\
\#8 & $1.0^{+0.8}_{-0.4}$ & 104 & $0.3^{+0.2}_{-0.1}$ & 81 \\
\hline
mean & $1.7^{+1.9}_{-0.8}$ & 90 & $0.6^{+0.5}_{-0.2}$ & 72 \\
\hline
\label{ResTabA}
\end{tabular}
\end{table}

\begin{table}
\centering
\caption{
	Same as Table \ref{ResTabA} but this time using the MACHO
collaboration's efficiency function B.  }
\begin{tabular}{lcccc}
\hline
 Halo & $\langle\tau_{\rm cone}\rangle$   & $\hat{t}_{\rm exp}$ & $\langle\Gamma_{\rm cone}\rangle$  & $\hat{t}_{\rm exp}$ \\ 
      & $10^{-7}$		 & days       & $10^{-6}$        & days     \\
      & & & events & \\
      & & & star$^{-1}$ yr$^{-1}$ & \\
\hline
\hline
\#1 & $2.7^{+2.9}_{-1.3}$ & 84 & $0.9^{+0.8}_{-0.4}$ & 66 \\
\#2 & $2.9^{+3.6}_{-1.3}$ & 84 & $1.0^{+1.0}_{-0.4}$ & 66 \\
\#3 & $2.1^{+1.9}_{-0.9}$ & 84 & $0.7^{+0.5}_{-0.3}$ & 64 \\
\#4 & $2.4^{+2.7}_{-1.2}$ & 86 & $0.8^{+0.7}_{-0.3}$ & 74 \\
\#5 & $2.0^{+1.6}_{-1.1}$ & 94 & $0.6^{+0.4}_{-0.3}$ & 69 \\
\#6 & $2.4^{+2.5}_{-0.9}$ & 99 & $0.8^{+0.6}_{-0.3}$ & 71 \\
\#7 & $2.3^{+3.2}_{-1.2}$ & 89 & $0.7^{+0.8}_{-0.3}$ & 61 \\
\#8 & $1.3^{+1.0}_{-0.5}$ & 104 & $0.4^{+0.3}_{-0.1}$ & 81 \\
\hline
mean & $2.3^{+2.4}_{-1.0}$ & 90 & $0.7^{+0.7}_{-0.3}$ & 69 \\
\hline
\label{ResTabB}
\end{tabular}
\end{table}

\section{Discussion}

The benefits of using $N$-body haloes instead of analytical models are: (i) We
do not have to make any initial assumptions about the velocity distribution of
the matter (other than limiting the circular velocity to 220 km/s), whereas
analytical models have to adopt a Maxwellian distribution. (ii) We do not have
to make any initial assumptions about the shape of the halo, whereas analytical
models are always educated guesses about the shape in form of some given
parameters (e.g. triaxiality, density profile, etc.).

\cite{wd} used a so called microlensing tube to get better
number statistics for their event rates, and as they note, this ``distorts the
geometry of a realistic microlensing experiment''. However, our solution (the
introduction of subparticles in accordance to the TSC mass assigment scheme) to
the insufficient mass resolution of the cosmological simulation {\it preserves}
the geometry.

The downsides of using $N$-body haloes are: (i) We assume that the spatial and
velocity distribution of MACHOs follows dark matter. (ii) The clusters we use
are not as old as the Milky Way.

Our models are based upon pure dark matter simulations which may not be
appropriate if a significant fraction of the dark matter is composed of MACHOs,
since they are composed of baryons, and baryonic physics has been explicitly
ignored! We thus implicitly assume that the spatial and velocity distribution
of MACHOs follows the respective distributions of the underlying dark
matter. Based upon these assumptions dark satellites comprised of MACHOs would
be detectable through excess optical depth values and event duration anomalies
in microlensing experiments. Simulations suggest that the majority of these
dark satellites can be located as far as 400 kpc from the halo centre. Thus,
multiple background sources at distances {\it over} 400 kpc would be needed to
detect possible dark subhaloes in a Milky Way sized dark halo. For example,
POINT-AGAPE (\citealt{belo}, \citealt{point-agape}) is a survey that is in principle able to detect even a
dark MACHO satellite, on top of ``free'' MACHOs in the dark halo, because its
source, M31, is distant enough.

From the results, we can see that Halo \#8 is too young to be used as a Milky
Way dark halo model, but all the other haloes seem to behave well.
The fact that Halo \#8 consists of three merging smaller haloes
explains the peculiar values it produces. The other haloes are not experiencing
any violent dynamical changes --- a requirement we would expect a model of the
Milky Way dark halo to fulfill.

Obviously, our MACHO mass function is a $\delta$-function. A more complex function
would force us to calculate event durations and rates for individual MACHOs instead of subparticles.
In this sense, subparticle values are only averages and some of the variation in the
observables is lost.

We chose not to use a complex MACHO mass function in this study because we wanted
to concentrate on the halo structure effects. The variations of $\Gamma$ and $\hat{t}$ would be larger
if we would use a MACHO mass function covering a large range of mass values. Thus, our results are as
conservative as possible, and in real experiments even larger variations of event rates and durations
could be expected. 

We recognize the fact that in the MACHO-lens scenario some of the lensing events towards the LMC could well be assigned to MACHOs within the dark halo of the LMC itself. We have excluded the LMC's
MACHO population in this study, but intend to investigate
the characteristics and implications of such lens population in a separate, future study.
Our plan is to model the dark matter halo of the LMC with some of the $M \sim 10^{9}$ \Msun{} subhalos found near the cores of our $N$-body haloes.

\section{Conclusions}

The purpose of this paper is to investigate the main microlensing characteristics of
$N$-body dark matter haloes, extracted from cosmological simulations and
downscaled in size and mass to represent the dark halo of the Milky Way. 

We argue that analytical halo models are too simplified in a number of ways in
the case of microlensing where internal structures (in density or velocity
distributions) or irregular halo shapes can alter the observed values
significantly.

We find that in general observables behave as expected from analytical
models, resulting in fairly consistent values with seven out of the eight haloes
with the ``outsider'' being exceptionally young (3.42 Gyr) and undergoing a
major merger of three smaller entties. As a result, this particular halo shows
irregular behaviour in all tests and can not be considered as a valid model for
the Milky Way dark halo.

When individual experiments are examined in detail, we find that triaxiality
and substructures can have a large effect on $\tau$ and $\Gamma$. In some
haloes, triaxiality can change the observed values by a factor as large as
three. Substructure within haloes can also change $\tau$ and $\Gamma$ by a
factor of two and furthermore reshape the event duration distribution notably.

We also use our simulated values together with the MACHO collaboration's observations
to find a preferred halo MACHO fraction ($f$) and individual MACHO mass ($m_{\rm M}$).
Our results are similar to the MACHO collaboration's own analysis where they used several 
different analytical models. In our analysis of $f$ and $m_{\rm M}$, the scatter between different microlensing experiments is mainly due to triaxiality and no clear signs of frequent substructure
signals can be seen.

\section*{Acknowledgments} 

The cosmological simulations used in this paper were carried out on the Beowulf cluster
at the Centre for Astrophysics \& Supercomputing, Swinburne University.

AK acknowledges funding through the Emmy Noether Programme by the DFG (KN 722/1).

The financial support of the Australian Research Council; the Jenny and
Antti Wihuri Foundation; the Magnus Ehrnrooth Foundation; 
Finnish Academy of Science and Letters, Vilho, Yrj\"o and Kalle V\"ais\"al\"a Foundation 
and the Academy of Finland are gratefully acknowledged.

JH gratefully acknowledges the hospitality
of Swinburne University, where the inital part of this work was performed, 
with especial thanks to BG, and
the Astrophysical Institute of Potsdam with especial thanks to AK.  


\onecolumn

\end{document}